Dielectric Properties of Disordered $A_6B_2O_{17}$ (A = Zr; B = Nb, Ta) Phases

R. Jackson Spurling[1,]*, Saeed S.I. Almishal[1], Joseph Casamento[1,2], John Hayden[1], Ryan Spangler[1], Michael Marakovits[3], Arafat Hossain[4], Michael Lanagan[3], & Jon-Paul Maria[1,]*

[1]Department of Materials Science Engineering, Pennsylvania State University

[2]Department of Materials Science Engineering, Massachusetts Institute of Technology

[3]Department of Engineering Science and Mechanics, Pennsylvania State University

[4]Department of Electrical Engineering, Pennsylvania State University

*E-mail: rjs7012@psu.edu, jpm133@psu.edu

Abstract: We report on the structure and dielectric properties of ternary $A_6B_2O_{17}$ (A = Zr; B = Nb, Ta) thin films and ceramics. Thin films are produced via sputter deposition from dense, phase-homogenous bulk ceramic targets, which are synthesized through a reactive sintering process at 1500 °C. Crystal structure, microstructure, chemistry and dielectric properties are characterized by X-ray diffraction and reflectivity, atomic force microscopy, X-ray photoelectron spectroscopy, and capacitance analysis respectively. We observe relative permittivities approaching 60 and loss tangents $< 1 \cdot 10^{-2}$ across the $10^3$-$10^5$ Hz frequency range in the $Zr_6Nb_2O_{17}$ and $Zr_6Ta_2O_{17}$ phases. These observations create an opportunity space for this novel class of disordered oxide electroceramics.

1. Introduction

Recent reports exploring chemical diversity and disorder in oxide materials [1] have opened new avenues for designing materials with tailorable functionalities [2–8]. A structural family of particular interest is $A_6B_2O_{17}$ (A = Zr, Hf; B = Nb, Ta) [9–11]. While the structure and thermodynamic behavior of these phases are now well-understood, comparatively few reports address property measurements. This manuscript provides an initial survey of $A_6B_2O_{17}$ low field dielectric properties as a function of frequency for thin films and ceramics.

Galy and Roth provided the first structure solution for the $Zr_6Nb_2O_{17}$ formulation in 1973 [12] (modelled in Figure 1, after Spurling, et al [11]). More recent reports by McCormack and Kiven [9], Voskanyan, et al [10], and Spurling, et al [11] show that the different ternary and quinary permutations of the $A_6B_2O_{17}$ phase are isomorphous (orthorhombic, space group *Ima2*) and contain substantial cation sublattice disorder character; furthermore, they discuss the entropic contributions to thermodynamic phase stability and enhanced solid solubility. This family of phases is unique due to its tolerance for chemical complexity and the variety of available local coordination environments: 6-, 7-, and 8-fold coordinations are present [9,12]. Given the substantial configurational entropy arising from both site and chemical complexity, the $A_6B_2O_{17}$ family is an attractive system for exploring the interplay between structural disorder and properties. Recent reports have investigated high-temperature chemical and mass transport [13,14] and mechanical properties [15] of $A_6B_2O_{17}$ phases; however, electronic properties have hitherto received limited attention. While the dielectric properties of the endmember binary oxide species are well known [16], similar reports for $A_6B_2O_{17}$ permutations are not available in the open literature. Moreover, studies of the $A_6B_2O_{17}$ phases have predominantly focused on bulk ceramic synthesis [9–12,14,17,18] and consequently comparatively little is known about their overall physical properties or the opportunities they may present as thin layers. We note the growing interest in the published literature for understanding synthesis and properties of disordered, multi-constituent oxides [19] and emerging electronic materials [20,21] in the thin film regime; these developments collectively provide the impetus for property studies in $A_6B_2O_{17}$. The $A_6B_2O_{17}$ presents a unique opportunity for dielectric tunability design using elemental selection and disorder [7]. Here, we explore the thin film growth and dielectric characteristics of two $A_6B_2O_{17}$ permutations: $Zr_6Nb_2O_{17}$ and $Zr_6Ta_2O_{17}$.

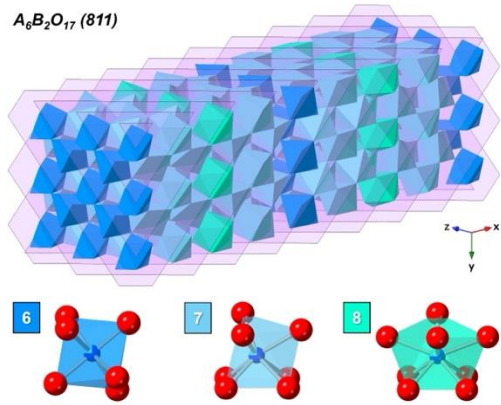

***Fig. 1.*** *Structural model of the $A_6B_2O_{17}$ unit cell and local cation coordination environments with the (811) crystallographic planes (most prominent in bulk and thin film diffraction) superimposed, per the structure solution provided by Galy and Roth [11] and McCormack and Kriven [8] (model rendered using CrystalMaker® software suite).*

The $Zr_6Nb_2O_{17}$ and $Zr_6Ta_2O_{17}$ permutations of the $A_6B_2O_{17}$ phases are chosen for the present study because they are readily densified through a conventional reactive sintering procedure at atmospheric pressure. We note that the $Hf_6Ta_2O_{17}$ isomorph can be difficult to densify, however densification has been reported under higher temperature sintering conditions [14]. The present study investigates sensitivities of crystallinity and dielectric behavior to oxygen partial pressure and total deposition pressure, motivated by existing reports of such sensititvities in sputter deposited endmember binary oxide films [22,23].

## 2. Experimental Methods

Thin films are synthesized via sputter deposition from densified bulk ceramic targets. Bulk ceramic sputter targets are prepared following a similar reactive sintering procedure as outlined in the literature [11]. Briefly, stoichiometric amounts of as-received oxide powders ($ZrO_2$, TOSOH, 99.87%; $Nb_2O_5$, Sigma Aldrich, 99.99%; $Ta_2O_5$, Sigma Aldrich, 99.5%) are ball milled in methanol (Fisher Scientific, Grade: Certified ACS Reagent) using yttrium-stabilized zirconia media for 30 hours. Powders are air-dried at 100 °C to drive off methanol for at least 2 hours, followed by uniaxial pressing to form ~10g pellets in a 1.25 inch cylindrical steel die (shrinkage and porosity reduction during sintering will ensure that the resulting targets have a diameter of ~1 inch). Green bodies are air sintered at T = 1500 °C for 12 hours to obtain reacted, densified ceramics. Volume density measurements are made of the roughly cylindrical sintered targets to confirm sufficient mechanical robustness and sputter yield consistency; we establish a general density threshold of > 90% $\rho_{Theoretical}$.

Thin film synthesis is conducted via radio frequency (RF) magnetron sputter deposition from the 1-inch ceramic $Zr_6Nb_2O_{17}$ and $Zr_6Ta_2O_{17}$ targets mounted in AJA 1" Stiletto sources. Pre-electroded $Pt/Ti/SiO_2/Si$ substrates with 111 oriented Pt are used. For all depositions reported here, only the oxygen content of the plasma (measured as a percentage of the total volumetric gas flux into the chamber) and the total deposition pressure are varied; the substrate stage temperature (350 °C), gun power (30 W), target-substrate throw distance (~ 1 inch), and total volumetric gas flux (20 sccm) are held constant. Gas flow is controlled with separate electronic flow meters (Alicat Scientific, 0.1 sccm control). The base pressure of the chamber prior to the start of all depositions is below $1 \cdot 10^{-6}$ Torr; the deposition rate in this configuration is measured as ~0.5 nm/min. X-ray diffraction (XRD) and reflectivity (XRR) measurements (Panalytical Empyrean, Bragg-Brentano geometry, Cu k$\alpha$ $\lambda$ = 1.54 Å) are used to confirm crystallinity and thickness/roughness of as-deposited films while atomic force microscopy (AFM) is used to characterize film surface morphologies and roughness. After structural characterization, Pt is sputtered in a pure Ar environment onto masked samples to form top electrodes; the $Pt/Ti/SiO_2/Si$ substrate acts as a common ground plane.

X-ray photoelectron spectroscopy (XPS) experiments are performed using a Physical Electronics VersaProbe III instrument equipped with a monochromatic Al k$\alpha$ x-ray source (h$\nu$ = 1,486.6 eV) and a concentric hemispherical analyzer. Charge neutralization is performed using both low energy electrons (<5 eV) and argon ions. The binding energy axis is calibrated using sputter cleaned Cu (Cu $2p_{3/2}$ = 932.62 eV, Cu $3p_{3/2}$ = 75.1 eV) and Au foils (Au $4f_{7/2}$ = 83.96 eV). Peaks are charge-referenced to $CH_x$ band in the carbon 1s spectra at 284.8 eV. Measurements are made at a takeoff angle of 85° with respect to the sample surface plane to maximize sampling depth. This results in a typical sampling depth of 6-8 nm (95% of the signal originated from this depth or shallower). The analysis size is 50-200µm in diameter depending on the sample.

Capacitance and loss tangent measurements of thin films are conducted using a Keithley 4200 parameter analyzer. Relative permittivities are calculated assuming a basic parallel plate capacitor, with electrode area measurements collected individually using scanning electron microscopy; film thickness measurements are made using XRR. Capacitance and loss are measured as a function of frequency for the $10^3$-$10^5$ Hz range. Microwave frequency dielectric measurements are performed concurrently on bulk ceramic samples at room temperature following the Hakki-Coleman approach [24].

3. Results and Discussion

X-ray diffraction for ceramic targets of $Zr_6Nb_2O_{17}$ and $Zr_6Ta_2O_{17}$ bulk phases after 12 hours at 1500 °C are shown in Fig. (2a), sharp peaks with relative intensities consistent with ICCD reference data suggest that reactions are proceeding to completion in both. Scanning electron microscopy for the $Zr_6Ta_2O_{17}$ target surface is shown in Fig. 2(b); residual porosity is modest with no obvious microstructural signs for phase segregation.

The initial thin film experiment compared film crystal structure and electrical properties (at constant deposition temperature and pressure) as a function of Ar:$O_2$ ratio. Figure 3 shows a representative set of diffraction patterns for $Zr_6Ta_2O_{17}$ grown in this series that focus on the 811 reflection so as to highlight crystallinity trends. For reference, a complete pattern is provided in Fig. 2(a) allowing comparison to the bulk data. XRD shows a strong preference for growth in the 811 orientation (~ 30° 2θ), which is also the most prominent diffraction peak in the bulk ceramic diffraction pattern (and calculated pattern) and which comprises the oxygen close packing planes. We note the relatively narrow $O_2$ content window (measured as a function of volume percentage with a balance of Ar) that suppports crystalline $A_6B_2O_{17}$ film growth. Films grown in a pure Ar environment are amorphous, while 10% $O_2$ represents a general upper bound; films sputtered in higher oxygen concentrations exhibit poor lateral uniformity, reduced crystallinity, and lower growth rates, these effects are associated with anion re-sputtering which authors have previously noted in comparable systems [25–27]. The $Zr_6Nb_2O_{17}$ system exhibits similar trends, with a comparable overall functional dependency, albeit with a higher oxygen threshold for producing crystalline thin films.

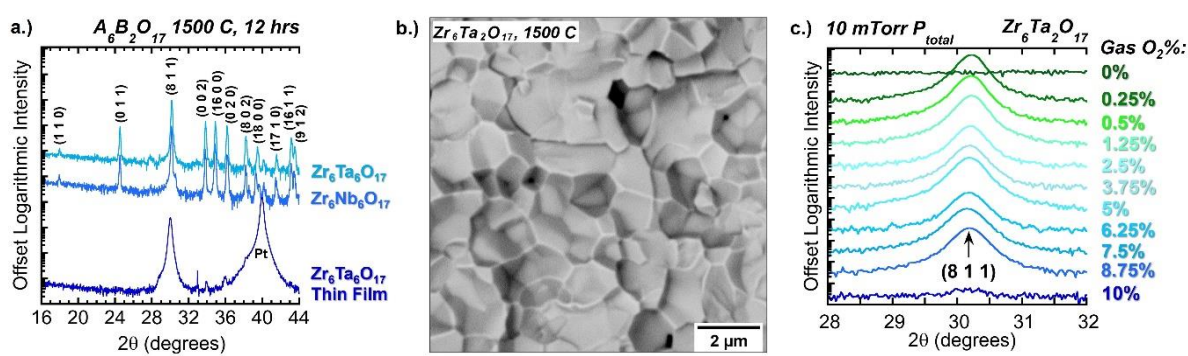

*Fig. 2. (a) XRD theta-twotheta scans for $Zr_6Nb_2O_{17}$ and $Zr_6Ta_2O_{17}$ sintered ceramic targets and a Zr6Nb2O17 thin film (b) and backscatter electron micrographs of a $Zr_6Ta_2O_{17}$ target; and (c) XRD scans for $Zr_6Ta_2O_{17}$ samples grown as a function of oxygen partial pressure.*

Figs. 3(a-d) show the results for capacitor testing in both materials for all preparation conditions. For the niobate system, the lowest oxygen partial pressures produce dielectric properties that show strong dispersion and high loss, likely associated with incompletely oxidized Nb. Increasing the oxygen fraction reduces the loss dramatically, with a peak relative permittivity value occurring at 2.5% oxygen and then falling with additional concentrations. The peak value at 2.5% is associated with enough oxygen present to fully oxidize the incoming cations and not so much as to induce bombardment damage, which can strongly change film density and microstructure. Loss tangent values for all samples at higher oxygen partial pressures are in the range near 0.03. For the tantalate system, the overall behavior is comparable. Pure argon deposition conditions produce lossy and dispersive dielectric responses with a general trend of higher relative permittivity for more oxygen rich conditions. In this case, the peak relative permittivity occurs for 5% oxygen with a strong reduction at 10% oxygen. Again, the most interesting electrical properties occur at an intermediate oxygen content condition set that likely balances maximum oxidation without deleterious bombardment.

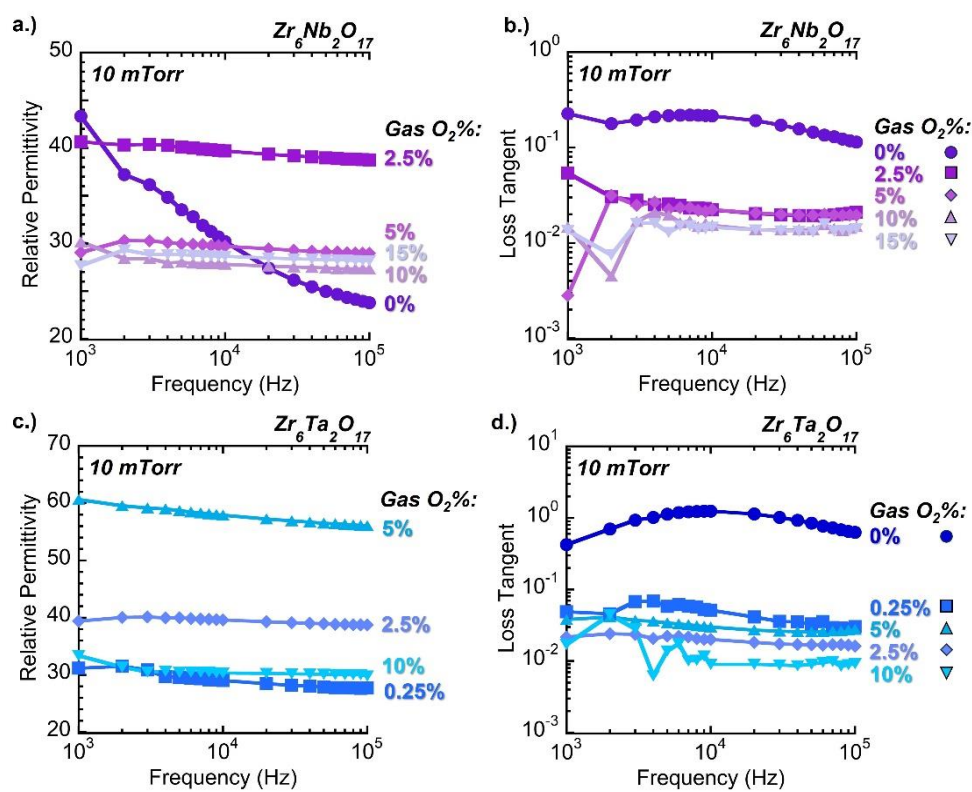

*Fig. 3. Relative permittivitiy and loss tangent measurements for sputter-deposited $Zr_6Nb_2O_{17}$ (a & b) and $Zr_6Ta_2O_{17}$ (c & d) samples grown under variable oxygen partial pressure.*

The strongest effects observed for this initial data set suggest that oxidation power of the plasma and oxygen impingement has the strongest effect on structure and properties with a correct balance producing the best dielectric properties. Enough bombardment is needed to promote surface mobility, crystallinity and density – this fraction likely coming from Ar and the metal adatoms – but not so much as to cause damage – this fraction coming from negative anions. As such, we chose a second experiment where the oxygen fraction is fixed at 2.5%, but total pressure is varied between 1.5 and 20 mTorr. In this case, more total pressure will produce a more oxidizing plasma, but more total pressure will also thermalize the overall average plasma kinetic energy. For this experiment, we introduce the electrical properties first in Fig. 4 which shows relative permittivity and loss tangents for all samples.

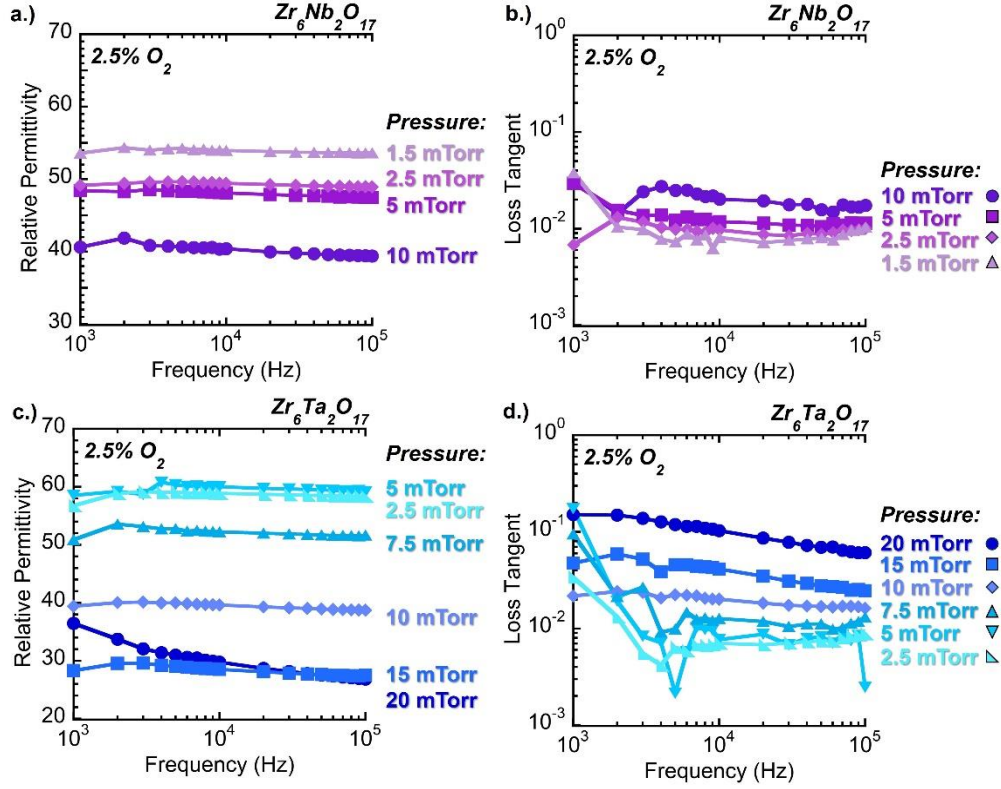

*Fig. 4. Relative permittivitiy and loss tangent measurements for sputter-deposited $Zr_6Nb_2O_{17}$ (a & b) and $Zr_6Ta_2O_{17}$ (c & d) samples grown under variable total pressure conditions.*

For this experiment, in both compositions, the trends were monotonic and easier to interpret. Generally, reducing total pressure increases relative permittivity and lowers loss when the oxygen fraction is fixed at 2.5%. More importantly, samples with the highest relative permittivity also had the lowest loss and the smallest frequency dispersion. For the highest relative permittivity samples, loss tangent values were in the mid to high $10^{-3}$ range. Perhaps more interestingly, tanδ values are consistent with literature reports of sputtered thin films of endmember binary oxides [28,29] but the relative permittivity values are approximately a factor of two larger which suggests that longer range contributions to the polarizability (beyond electronic and ionic contributions) are present. Hakki-Coleman measurements of bulk ceramics confirm a relative permittivity of ~60 for both $Zr_6Nb_2O_{17}$ and $Zr_6Ta_2O_{17}$ and loss tangents well below $1·10^{-2}$ tanδ in the GHz frequency range, corroborating the thin film study presented here. Collectively, these relative permittivity values and generally low loss tangents imply that these ternary $A_6B_2O_{17}$ phases may be interesting in dielectric material applications and merit further investigation.

We attribute trends in relative permittivity and loss to differences in film density and microstructure, and differences in charge defect concentrations. While these can be challenging to quantify in thin layers with small grains produced at modest temperatures, we attempt to do so with a combination of X-ray diffraction, X-ray reflectometry, and X-ray photoelectron spectroscopy. For X-ray diffraction and reflectometry, we highlight the tantalate system. Fig. 5 summarizes these results.

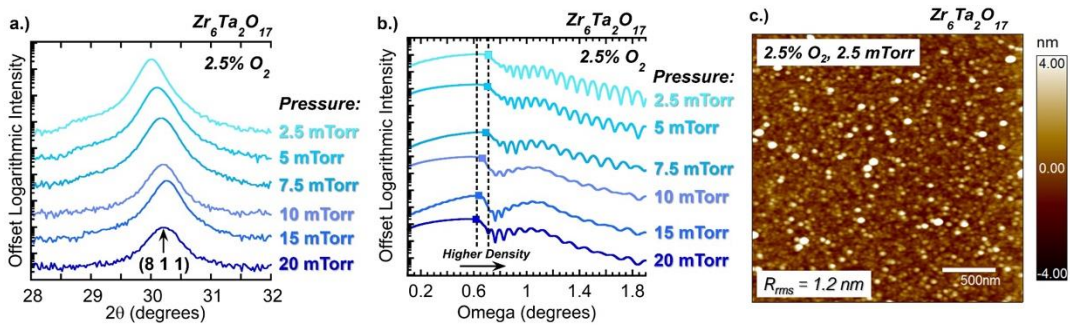

*Fig. 5.* XRD (a) scans for $Zr_6Ta_2O_{17}$ thin films demonstrate a modest lattice parameter shift and increase in crystallinity with decreasing total deposition pressure. XRR (b) scans of the same films show a critical angle shift indicating increasing density with decreasing deposition pressure (all films are between ~60-90 nm thick as obtained via XRR fitting). A representative AFM scan (c) of a film grown at low pressure demonstrates smooth growth and controlled roughness.

Fig. 5(a) again highlights the $A_6B_2O_{17}$ (811) reflection as a function of pressure. The peak intensity increases substantially with falling pressure and the predominantly out-of-plane interplanar spacing shifts to larger values. For the lower pressure condition, the (811) d-spacing is consistent with the bulk value. Fig. 5(b) shows XRR scans for each film; again, a general trend persists where best fit models suggest smoother films with higher density at lower pressures. The AFM image in Fig. 5(c) shows the physical topography, very small grains (~50 nm) and an RMS roughness below 2 nm – both values are consistent with expectations for refractory oxides prepared at low temperature. This entire data set is also consistent with the dielectric properties; in most general terms, high relative permittivity values and low loss tangents are anticipated when the film microstructure is dense, when the lattice constants are close to bulk values, when surfaces are smooth, and when cations are in the fully oxidized state.

XPS analysis is used to explore the final aspect of stoichiometry and cation valence, which can provide insights into possible defect population types that influence the dielectric response. XPS results are summarized in Fig. 6 for tantalate films prepared at low and high total pressures and constant oxygen fraction. Relative peak intensities indicate uniform cation stoichiometry to within instrumental limits, however, there is a noticeable shift (Fig. 6(a)) in the Ta 4f peak to a lower binding energy for material prepared at higher total pressure. It is interesting to consider that lower binding energies are consistent with a smaller chemical shift and an underoxidized state. Peak fitting of the 4f peaks indicates 100% $Ta^{5+}$ for 1.5 mTorr material and 35% $Ta^{4+}$/65% $Ta^{5+}$ for 10 mTorr material, even though there is more total oxygen present at high total pressure. This could be the result of lower kinetic bombardment at the high pressure condition accompanied by a larger ambient argon concentration (and therefore a more reducing atmosphere). We note that underoxidation of *B* species cations has been proposed as a defect mechanism to compensate for oxygen vacancies and which would produce charge carriers [30]; the present XPS results are consistent with that proposal. Regardless of the physical mechanism controlling cation valence, the XPS are consistent with electrical data – samples with the cations in their most stable oxidation state for a given structure should have fewer compensating charge defects that typically contribute to a larger loss tangent.

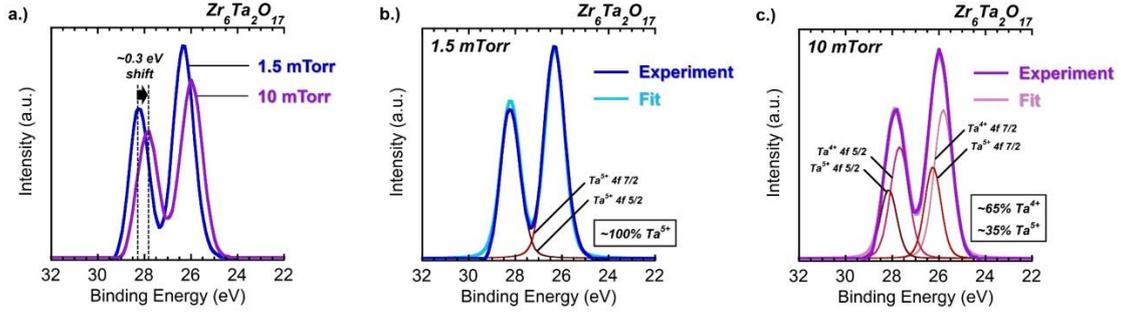

*Fig. 6.* (a) XPS scans for representative $Zr_6Ta_2O_{17}$ thin films grown as a function of pressure demonstrate a ~0.3 eV chemical shift; (b) subsequent fitting of samples grown at 1.5 mTorr; and (c) 10 mTorr demonstrating a shift from $Ta^{4+}$ to $Ta^{5+}$ states at low pressure conditions.

This final result is somewhat unexpected, as changing deposition conditions to include more oxygen will generally produce higher cation oxidation states; however, the opposite is observed presently. We note that previous literature reports for endmember oxide sputtering, such as $ZrO_2$, are only able to produce crystalline films at high oxygen partial pressures, with amorphous films resulting from low oxygen partial pressure conditions [22]. While we cannot definitively isolate the mechanism responsible for this apparent disparity, we speculate that in the present case microstructure prevails, and that the lower pressure conditions promote more favorable bombardment during sputtering. These favorable bombardment conditions will produce better crystals, denser films, and smoother surfaces, with fewer defects that either include or are compensated by lower valent Ta. This interplay between oxygen partial pressure and bombardment conditions could highlight particular sensitivities in the $A_6B_2O_{17}$ sputtering process. We note that the connection between sputtering conditions and defect chemistry control is central to ongoing disordered oxide materials research [19]. This presents a natural opportunity for future studies of defect chemistry resulting from processing changes in $A_6B_2O_{17}$. Regardless of the exact processing-based mechanism governing the film defects and microstructure under variable oxygen partial pressure presented here, the changes in density and crystallinity are significant and provide context for the electrical data.

4. Conclusions

In this paper we use a series of characterization techniques to study the structure and dielectric properties of sputter deposited $Zr_6Nb_2O_{17}$ and $Zr_6Ta_2O_{17}$ thin films. We report a strong dependence of crystallinity, roughness, and dielectric properties (including relative permittivity and loss) on the plasma composition and total pressure during sputtering. These materials have high relative permittivities in comparison to endmember binary oxides while also retaining low loss tangent values across the $10^3$-$10^5$ Hz frequency range, making them attractive for further property exploration and potential usage in dielectric material applications. This report provides further motivation for studying the interplay between structure, processing, and properties of disordered oxides with the potential for significant future impact on application-driven electroceramics engineering design.

Acknowledgements:

This material is based upon work supported by the National Science Foundation, as part of the Center for Dielectrics and Piezoelectrics under Grant Nos. IIP-1841453 and IIP1841466. RJS was supported by the National Science Foundation Graduate Research Fellowship Program under Grant No. DGE1255832. Any opinions, findings, and conclusions or recommendations expressed in this material are those of the authors and do not necessarily reflect the views of the National Science Foundation. SSIA and JPM gratefully acknowledge support from NSF MRSEC DMR-2011839. The co-authors acknowledge use of the Penn State Materials Characterization Lab, specifically the XPS instrumentation; the authors would also like to acknowledge Jeff Shallenberger for helpful discussions related to XPS data analysis and acquisition.